\documentclass[sigconf,authorversion,nonacm]{acmart}
\def\BibTeX{{\rm B\kern-.05em{\sc i\kern-.025em b}\kern-.08emT\kern-.1667em\lower.7ex\hbox{E}\kern-.125emX}}

\newcommand{\nop}[1]{}
\DeclareMathOperator*{\argmin}{argmin}

\DeclareMathOperator*{\ra}{\longrightarrow}

\begin{document}

%

\title[KDD Research Paper]{Meta-Path-based Probabilistic Soft Logic for Drug-Target Interaction Prediction}

%

\author{Shengming Zhang}
\email{shengming.zhang@rutgers.edu}
\affiliation{%
  \institution{Rutgers University}
  \city{Newark}
  \state{New Jersey}
  \country{USA}
}

\author{Yizhou Sun}

\email{yzsun@cs.ucla.edu}
\affiliation{%
  \institution{University of California, Los Angeles}
  \city{Los Angeles}
  \state{California}
  \country{USA}
}

%

%
\begin{abstract}
\nop{
Drug-target interaction (DTI) prediction is an area of rising interest in the Bioinformatics world because of the high cost of time and money for chemical-based drug design. Most of the recently proposed methods use single drug-drug similarity and target-target similarity information for DTI prediction and are unable to take advantage of the multi-similarity information. Even for some proposed methods that could enumerate multi-similarity information, they lack the ability to take the rich topological information of the heterogeneous network into consideration, and the time consumption of these approaches are relatively high. We propose a network-based drug-target interaction prediction approach that applies summated meta-path and probabilistic soft logic (PSL) on a heterogeneous network which contains multiple sources of information, including drug-drug similarities, target-target similarities, drug-target interactions and other potential information. Our approach is based on the PSL graphical model and uses summated meta-path to reduce the number of rule instances of PSL. We compare our model against five methods, on three open-source datasets reported in three of them. The experimental results show that our approach outperforms all the five baselines in the AUPR score and AUC score.}
Drug-target interaction (DTI) prediction, which aims at predicting whether a drug will be bounded to a target, have received wide attention recently, with the goal to automate and accelerate the costly process of drug design. Most of the recently proposed methods use single drug-drug similarity and target-target similarity information for DTI prediction, which are unable to take advantage of the abundant information regarding various types of similarities between them. Very recently, some methods are proposed to leverage multi-similarity information, however, they still lack the ability to take into consideration the rich topological information of all sorts of knowledge bases where the drugs and targets reside in. More importantly, the time consumption of these approaches is very high, which prevents the usage of large-scale network information. We thus propose a network-based drug-target interaction prediction approach, which applies probabilistic soft logic (PSL) to meta-paths on a heterogeneous network that contains multiple sources of information, including drug-drug similarities, target-target similarities, drug-target interactions, and other potential information. Our approach is based on the PSL graphical model and uses meta-path counts instead of path instances to reduce the number of rule instances of PSL. We compare our model against five methods, on three open-source datasets. The experimental results show that our approach outperforms all the five baselines in terms of AUPR score and AUC score.
\end{abstract}

%
%


%

%

\maketitle

\section{Introduction}
New drug development is usually a very time consuming and expensive procedure. Recently, computer aided drug design has received wide attention, with the goal to accelerate drug design. Among them, the study of predicting unknown drug-target interactions based on existing domain-specific knowledge using mathematical models becomes an area of growing interest \cite{doi:10.1093/bib/bbv066}. By quantitatively expressing the similarity between drug-drug and target-target, one can find a mathematical relationship between drugs and targets, which could help to predict potential interactions between existing drugs and unknown targets, or vice versa \cite{doi:10.1021/ci100176x}.

There are several existing methods to model the drug-target interaction prediction \cite{doi:10.1093/bib/bbt056} task, most of which apply a network-based representation \cite{PMID:17921997}: \cite{6817596} constructs a bipartite interaction network which has two types of nodes: drug nodes and target nodes. There can be edges between two drug nodes or two target nodes, denoting as similarity information. Edges between drug nodes and target nodes represent interaction information. However, such bipartite interaction network constrains the type of nodes within two, and is unable to add additional measures.

In addition to a bipartite interaction network, \cite{chen2015drug} constructs a heterogeneous internet that could directly integrate richer domain-specific knowledge into the network, such as drug-drug/target-target interaction information, drug-cure-disease and disease-caused-by-target information etc. Instead of transferring these information into similarities using standard measures (such as the Jaccard and Spearman indexes), the DTI prediction task can be solved using the information directly extracted out of the heterogeneous network.

Using link prediction methods can predict potential interactions within a network, which is proposed both in \cite{Getoor:2005:LMS:1117454.1117456} and \cite{LU20111150}. However, some link prediction methods ignore the inner relationship between different semantic similarity information \cite{Fu2016} and other domain-specific knowledge, while some methods tend to take all the detail into consideration to achieve good results, but the time consumption becomes very heavy \cite{6817596,cichonska2018learning,luo2017network}.


In this paper we present a drug-target interaction prediction method based on \textit{probabilistic soft logic} (PSL) \cite{kimmig:probprog12}. We predict unknown drug-target interactions using multi-relational information and existing interactions of the network. In order to avoid grounding out all the rule instances that could significantly slow down the inference process like the original PSL model, We apply summated \textit{meta-path} \cite{Fu2016} which defines several semantic meta-paths and uses matrix multiplications to calculate the path counts, storing as commuting matrices. We apply a Bayesian probabilistic approach that transfers the path counts into probabilities, indicating probabilities for the body parts of PSL rules \cite{bac:jmlr17}, then apply the PSL model for the DTI prediction. Our summated meta-path PSL model outperforms all the five baselines \cite{6817596,Fu2016,cichonska2018learning,luo2017network} in both AUPR score and AUC score on three open-sourced datasets together with significant time-consumption reductions. 

In this paper, we define several semantic meta-paths and use matrix multiplications to generate commuting matrices corresponding to each semantic meta-path, which is similar to \cite{Fu2016}. Afterwards, we apply a Bayesian probabilistic approach that transfers the path-counts of the commuting matrices into probabilities. Then we could define several PSL rules. The first type of the PSL rules is: "Each semantic meta-path metric may imply potential interactions"; the second type of PSL rules is: "By default, drug and target does not interacts with each other". The first type of rules corresponds with the pre-defined semantic meta-paths, and the second type of rules is a negative prior. Different with the original PSL model, for each drug-target pair, we only have one rule instance within each PSL rule, since we applied summation using meta-path counts on all the existing rule instances. For each drug-target pair, we could treat the probabilities based on different semantic meta-paths as the body part of the PSL rule instances. For more details about the \textit{probabilistic soft logic} (PSL) please refer to \cite{bac:jmlr17,bach:uai13}. 

Our main contribution and novelty is that we figure out the shortcuts in both Meta-path method and PSL method and provide a perfect solution: Although the Meta-path approach proposes a robust path count topological feature, it does not give the feature a comprehensible meaning, only treating them as vector feature; the PSL model is a good probabilistic graphical model, but it tends to ground out every single rule instances and take them all into consideration, which results in heavy time consumption. By using the PSL model, the meta-path count topological feature can get a probabilistic comprehension, and by using path count summation strategy, the total number of rule instances for PSL significantly reduces so that it could be used for much larger datasets.

Based on the new settings, We implement a new DTI prediction framework, and compare our model with other five multi-similarity or internet-based approaches \cite{6817596,Fu2016,cichonska2018learning,luo2017network} on three open-sourced datasets used by \cite{6817596,Fu2016,luo2017network} (\cite{Fu2016} proposes two methods). The experimental results indicate that our model significantly reduces the running time as well as outperforming all five baseline models on all three datasets in AUC score and AUPR score.

\section{Related Work}
There are a number of Network structure-based methods for drug-target interaction predictions. \cite{cockell2010integrated} constructs a network structure \textit{metagraph} to organize drug, target, protein and genes showing their relationships. They point out that drugs with resemble structures can behave similarly in interactions. Based on a network structure, \cite{cheng2012prediction} treats the DTI prediction problem as an inference problem on a drug-target bipartite network which integrates chemical drug-drug similarities, sequencial target-target similarities and known drug-target interactions.

\cite{yamanishi2008prediction} integrates compound structure similarities, protein sequence similarities and several open-sourced drug-target interaction datasets in a network, generating the \textit{Gold standard datasets} which characterizes the drug-target interaction network into four classes: Enzyme, Ion channel, GPCR and Nuclear receptor. More recently, there are a number of matrix factorization drug-target interaction prediction methods \cite{liu2016neighborhood,7407341} which give state-of-the-art results on the golden standard dataset generated by \cite{yamanishi2008prediction}. 

In order to take multiple similarities and additional domain-specific measures into consideration, \cite{luo2017network} uses a dimensionality reduction scheme, diffusion component analysis (DCA), to obtain informative feature representations based on relational properties, association information and topological context of each drug and target within a heterogeneous network. After generating the feature representations, they use a learned projection matrix that best projects the drug feature into protein space so that the distance between the projected feature vectors from drug space to target space and the interacting proteins is minimized.



\cite{chen2012assessing} uses advanced topological features such as distance and number of shortest paths between node pairs for drug-target predictions. Furthermore, \cite{Fu2016} uses semantic meta-path topological features and apply SVM and Random Forest algorithm as classifier on a network integrated with multiple objects, including compound, protein, disease and gene etc.

\cite{cichonska2018learning} uses a kernel-based approach that first learns a corresponding weight for each similarity measure, then calculates a corresponding weight for each pairwise kernel. By detecting high ranking values within the weighted summed kernel matrix for the unknown drug-target pairs, they could be considered as new interactions. The dataset provided by this paper is a drug bioactivity prediction to cancer cells dataset, which labels are numerical bioactivity measures. As a result, it is not proper to be used as a comparison dataset for DTI prediction use.

Instead of using topological features, \cite{perlman2011combining} combines multiple sources of drug-drug similarity and target-target similarity data together into features, using logistic regression as classifier. \cite{6817596} extracts a subset of drugs and targets from the dataset created by \cite{perlman2011combining}, formulating the DTI prediction as an inference problem on a network and use \textit{probabilistic soft logic} (PSL) framework for inference. 


There are also non-similarity based approaches that takes pre-calculated drug and target features out of raw descriptors as the input of their models \cite{ezzat2016drug,ezzat2017drug}. Furthermore, \cite{wen2017deep} uses deep learning based approaches for DTI prediction that takes the raw drug and target descriptors as input and train their model among interactions of all the FDA approved drugs and targets. Since the input of datasets provided by these approaches are descriptors, the datasets are not proper to be used for comparison experiments using our multi-similarity and internet-based approaches.

\section{Preliminaries}
\subsection{Meta-path Count Topological Feature}
A semantic meta-path defines a certain type of paths linking the starting and ending objects. The total number of path instances of one semantic meta-path can be treated as a topological feature which evaluates the strength of associations between the starting and ending objects and is also called path count \cite{Fu2016}. In the DTI prediction task, a meta-path starts from a drug, and ends with a target, meaning there is one valid meta-path instance between the drug and target. 

For instance, we can define two types of semantic meta-paths:
$$Drug \ra\limits^{similar\ to} Drug \ra\limits^{interacts} Target \eqno{(A)}$$
$$Drug \ra\limits^{interacts} Target \ra\limits^{similar\ to} Target \eqno{(B)}$$
Meta-path (A) indicates that if a drug interacts with a target while this drug has similarity with another drug, then the other drug is also likely to interact with this target. Meta-path (B) has a similar interpretation.

For each drug-target interaction pair, we first calculate the corresponding path counts within the dataset under each semantic meta-path, then concatenate the numbers into a topological feature vector, each dimension denoting as the path count value of one semantic meta-path.

\subsection{Hinge-loss Markov Random Fields and Probabilistic Soft Logic}
Probabilistic soft logic (PSL) uses first-order logic syntax to form a hinge-loss Markov random fields (HL-MRFs) model. A hinge-loss Markov random fields model is defined over continuous variables which can naturally assemble probabilities and other real-valued attributes. By applying a maximum a posteriori (MAP) inference on a Hinge-loss Markov random fields, we can efficiently get exact inference result for all variables, as the MAP inference on a HL-MRFs model is a convex optimization problem \cite{bach:uai13}. A hinge-loss Markov random fields can be formulized as a log-concave joint probability density function:
$$P(Y|X) = \frac{1}{Z}exp(-\sum\limits^{M}\limits_{r=1}\lambda_r\phi_r(Y, X)) \eqno{(1)}$$

Y is the set of unknown variables and X is the set of observed values. $\lambda$ represents the set of weight parameters, Z is a normalization constant. The potential function $\phi_r(Y, X)$ for HL-MRFs is defined as:
$$\phi_r(Y, X) = (\max{(l_r(Y, X), 0)})^p \eqno{(2)}$$

$l_r$ is a linear function of Y and X and $p \in \{1,2\}$. The MPE inference is also equivalent to minimizing the convex energy.

In the PSL setting, all the grounded out rule instances which are associated with both known values, such as similarity information, and unknown variables, such as potential drug-target interactions, will be treated as terms in the potential function of a HL-MRFs model.

More in detail, we consider a general form of PSL rule:
$$w : P(A,B) \wedge Q(B,C) \rightarrow R(A,C) \eqno{(3)}$$

R (A, C) represents a continuous target variable, such as the probability of interactions between drug A and target C; P (A,B) and Q (B,C) represent observed values, such as the similarity between drug A and drug B and the probability of interaction between drug B and target C. The soft logic defines a relaxed convex representation of this logical implication:
$$\max\{P(A, B) + Q(B, C) - R(A, C) - 1, 0\} \eqno{(4)}$$

This continuous value can be treated as the \textit{distance to satisfaction} of the logical implication.

The MAP inference algorithm aims to minimize the energy of a hinge-loss Markov random fields, which is equivalent to minimize the total weighted \textit{distance to satisfaction} for all the grounded out rule instances. A full description of PSL is described in \cite{bach2015hinge}.

\section{Problem Definition}
We consider the problem of drug-target interaction problem as inferring new edges between drug nodes and target nodes on a heterogeneous network. Given a set of drugs D = \{$D_1,...,D_m$\} and a set of targets T = \{$T_1,...,T_n$\}, the total potential interactions between drugs and targets can be denoted as an interaction matrix $I_{m\times n}$, where $I_{i,j} = 1$ represents a positive interaction between drug $D_i$ and target $T_j$. In addition, a set of domain-specific measures are represented as edges between nodes in the network. For example, the multiple drug-drug similarities \{$S^1_d,...,S^{k_d}_d$\} can be treated as edges between drug nodes, and the disease-caused-by-target measure can be treated as edges between disease nodes and target nodes. Figure 1 shows a heterogeneous network based on the dataset provided by \cite{Fu2016}.

\begin{figure}[h!]
\centering\includegraphics[width=3.2in]{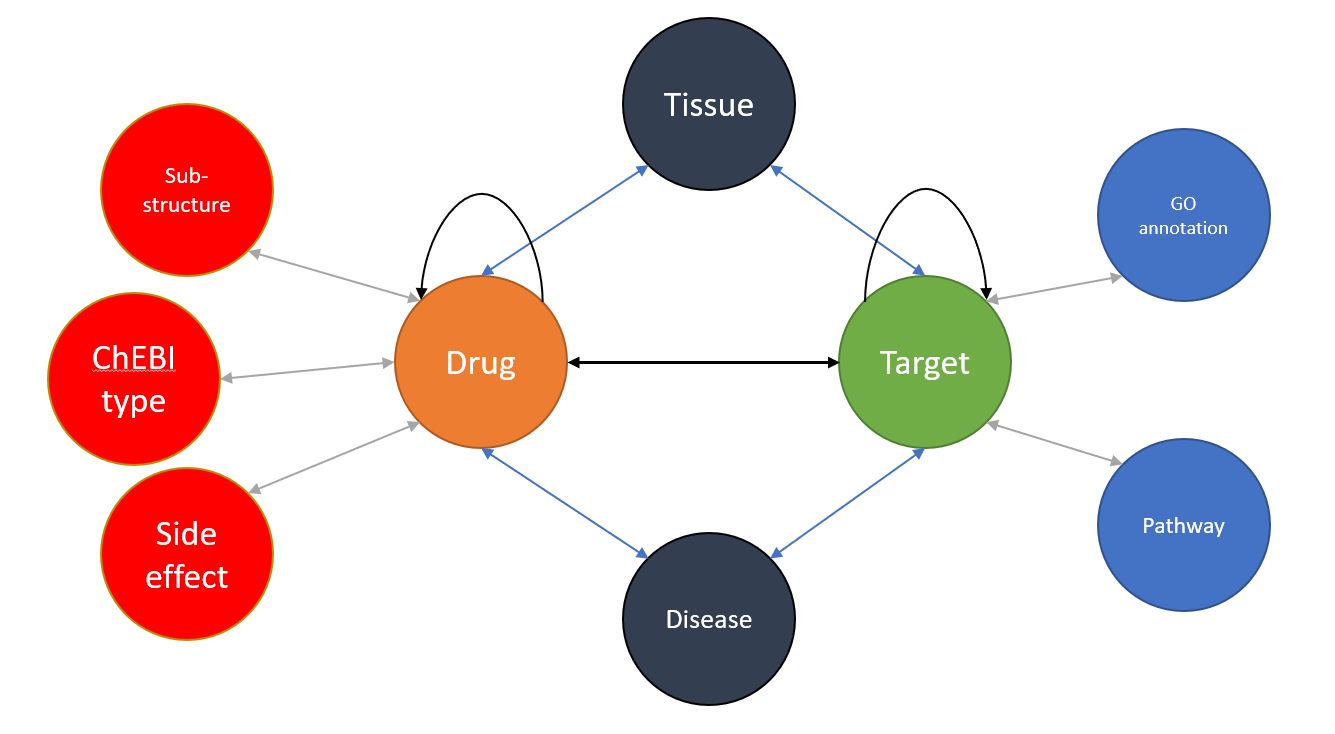}
\caption{A visual demonstration of the DTI prediction heterogeneous network.}\label{Figure 1} 
\end{figure}


The drug-target interaction prediction problem is to take use of all the information within the heterogeneous network to predict the unobserved interaction edges between drug nodes and target nodes.

\section{Summated Meta-path based Probabilistic Soft Logic (SMPSL)}

In this section, we are going to introduce our Summated Meta-path based Probabilistic Soft Logic (SMPSL) model. 

\subsection{Overview}

The principal aim for our method is to find an efficient and robust summation method onto the PSL rule instances, so that we could accelerate the inference, which is relatively slow in the original PSL model.

In a drug-target interaction network with m drugs and n targets, there are $mn$ possible interactions and if we consider all the similarities between each drug-drug and target-target pairs, the total count of meta-path instances can add up to $O(m^2n^2)$, which is very expensive for a large-scale network. Although \cite{6817596} applies a blocking method that finds the nearest k neighbors to pick the most similar drugs or targets and sets similarity values with farther neighbors to zero in order to reduce the total rule instances, the total number of rule instances can still be very big. Also, due to the fact that for some association measures, we only have binary value instead of continuous relevance, operating blocking method on these measures is unachievable. In order to reduce the number of rule instances, we propose our summated meta-path based probabilistic soft logic approach (SMPSL).

\subsection{Summated Meta-path based Probabilistic Soft Logic (SMPSL)}

\cite{Fu2016} introduces a topological feature, Meta-path count, that uses matrix multiplications to sum up the total count of all the meta-paths. Since both meta-path and PSL rule are association rules, we can take the same procedure to calculate a rule instance count.For instance, consider a drug-drug similarity based rule like this:

$$D1 \ra\limits^{Chemical} D2 \ra\limits^{Interacts} T1$$

It says that if drug D1 is similar in chemical fingerprints to drug D2, whil D2 is known to interact with target T1, then we could imply that drug D1 may also interacts with target T1. Figure 1 visualizes the process of generating a commuting matrix. 

\begin{figure}[h!]
\centering\includegraphics[width=3.2in]{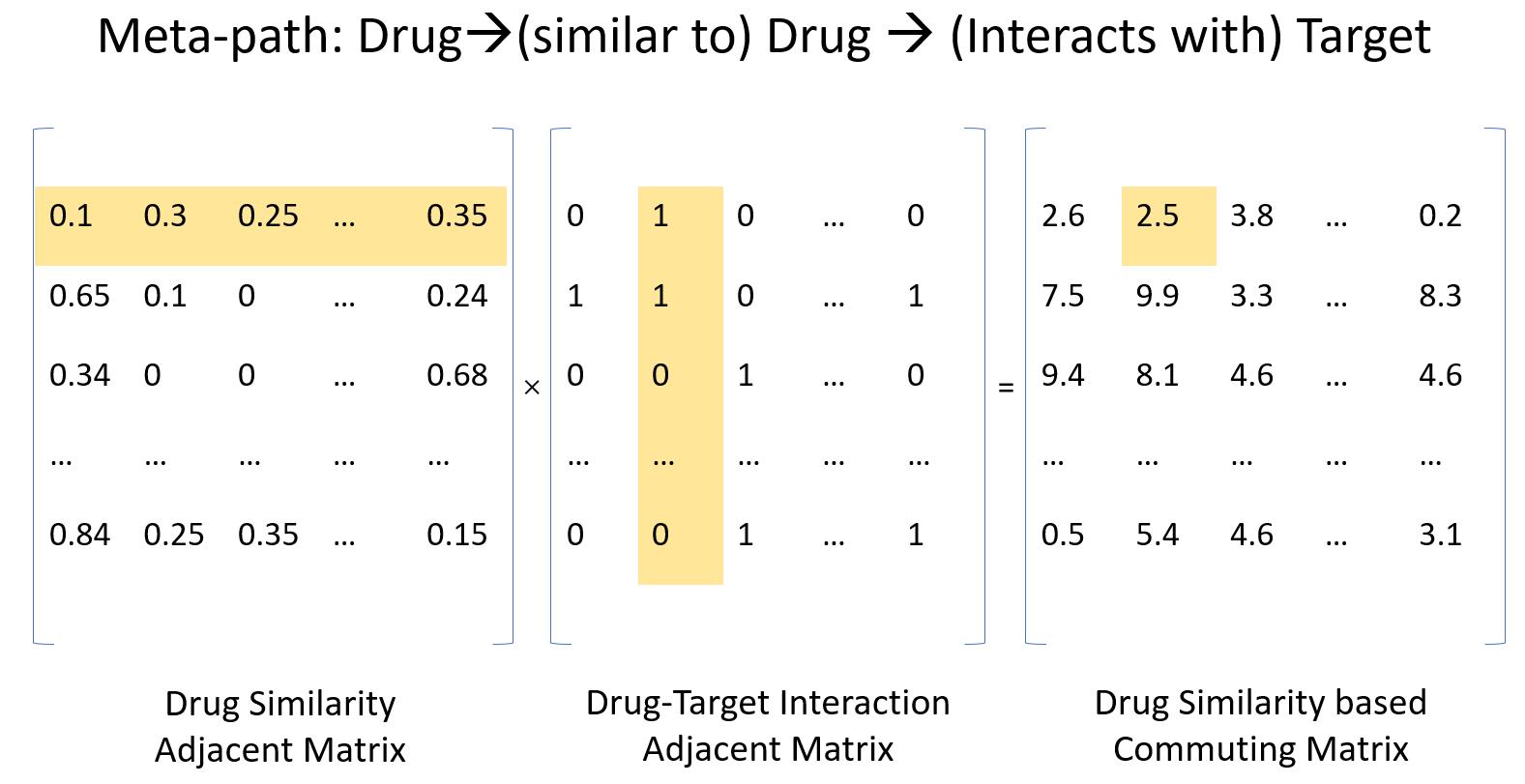}
\caption{Using matrix multiplications to generate a commuting matrix. Instead of using binary value for the similarity measure, we use exact similarities and generate a relaxed rule instance count.}\label{Figure 2} 
\end{figure}
Instead of using binary value for the similarity measure, we use exact similarities and generate a relaxed rule instance count. The matrix multiplication can be written as:

$${C}_{d}^1 = S_d^{1} \times I \eqno{(5)}$$

$C_d^1$ denotes to the commuting matrix of the \textit{chemical-based} drug similarity rule. $S_d^1$ is the \textit{chemical-based} similarity adjacent matrix, I is the drug-target interaction adjacent matrix. We could possibly consider multiple similarity measures and assign each rule with them by following the same procedure above. Similarly, if we consider another rule that introduces new types of nodes other than drug or target on a heterogeneous network:

$$Drug \ra\limits^{Treats} Disease \ra\limits^{Caused\ by} Target$$

If a drug is known to treat a certain type of disease, while a target is known as the cause of this disease, then we can imply that the drug may interacts with the target. The matrix multiplication form of calculating the commuting matrix for this meta-path can be written as:

$${C}_{Disease} = Treat_d \times Caused_t \eqno{(6)}$$

$C_{Disease}$ means the commuting matrix for disease-caused by rule, $Treat_d$ is the Drug-Disease association matrix, $Caused_t$ is the Disease-Target association matrix.

Although the value within commuting matrices can already represent the strength of drug-target pairs, we want to transfer these values into genuine probabilities so that they can fit in the probabilistic soft logic setting. We introduce a Bayesian probabilistic approach where the probability P that a drug-target interaction pair is positive interaction given commuting matrix value $C$ is defined as:

$$P( C ) = \frac{prior(1) \times \frac{c(C, 1)}{N(C, 1)}}{prior(1) \times \frac{c(C, 1)}{N(C, 1)} + prior(0) \times \frac{c(C, 0)}{N(C, 0)}} \eqno{(7)}$$

$c(C,label)$ is the number of drug-target pairs with known $label \in \{0, 1\}$ that has the same commuting matrix value, under the same PSL rule $C^l_d$. $N(C, label)$ denotes to the total number of non-zero valued drug-target pairs with known label k under the same PSL rule $C$. The prior probability for a label $prior(label)$ is defined as:

$$prior(label) = \frac{N(C, label)}{N(C, label) + N(C, 1-label)} \eqno{(8)}$$











After applying the Bayesian probabilistic approach, we generate only one rule instance for each pre-defined rules. We also include a negative prior rule which indicates that all drugs and targets tend not to interact with each other by default:

$$\neg Interacts(D, T) \eqno{(9)}$$

Based on the rules and probability obtained from the commuting matrices, we can write down our objective that minimizing the total energy on the Hinge-loss Markov Random Fields:

$$f(I^*) = \argmin\limits_{I}\sum\limits_{r \in R} w_{m+1}I_r^p + \sum\limits_{k=1}\limits^{m}w_k\max\{P_{r,k}-I_{r},0\}^p \eqno{(10)}$$

R is the union of all unknown drug-target interaction pairs. $I = (I_1,...,I_n) \in R$ represents the probability that unknown drug-target interaction pairs are positive. $w$ is a weight parameter, $p$ is an exponential parameter and in our experiment we take $p = 2$. Same as PSL model, given a settled weight parameters $w$, our objective aims to minimize the total \textit{distance to satisfaction} $d_r(I)^p$.

Due to the effect of summations, we significantly reduce the number of rule instances compared with original PSL framework, from over one million instances down to the number of defined PSL rules (meta-paths). Also, comparing with \cite{Fu2016}, we assign comprehensible propability to meta-path counts and use a probabilistic model to predict unknown links.

\textbf{Weight Learning}: In most cases, different rules may contribute unevenly, so adding a weight learning procedure before entering inference part is essential. We run a stochastic gradient descent algorithm on a portion of observed links to learn the weight parameters. The gradient of the log-likelihood with respect to the weight $w$ can be written as:

$$\frac{\partial}{\partial w} f(I) = -d_r(I)^p + E[d_r(I)^p] \eqno{(11)}$$

\section{Experimental Results}

\subsection{Datasets}

We use three datasets for our experiments. One is the dataset constructed and used by \cite{6817596,perlman2011combining}, which is a multi-similarity based dataset; another is used by \cite{Fu2016} and the other is used by \cite{luo2017network}, both of which incorporate additional domain-specific knowledge and form a heterogeneous network.

\subsubsection{Dataset I}

The dataset contains 315 drugs, 250 targets and 1306 known interactions. Besides, there are five drug-drug similarities and three target-target similarities within the dataset which are obtained from \cite{perlman2011combining}. The five drug-drug similarity measures are: Chemical-based, Ligand-based, Expression-based, Side-effect-based and Annotation-based. The three target-target similarity measures are: Sequence-based, Protein-protein interaction network-based and Gene Ontology-based. The drug-target interactions of this dataset are obtained from several open-source online databases organized by \cite{6817596}, including DrugBank \cite{wishart2007drugbank}, KEGG Drug \cite{kanehisa2009kegg}, Drug Combination database \cite{liu2009dcdb}, and Matador \cite{gunther2007supertarget}. A brief description of each similarity extraction is provided below:

\begin{itemize}
\item \textit{Chemical-based drug similarity:} \cite{perlman2011combining} use the chemical development kit \cite{steinbeck2006recent} to compute a hashed fingerprint for each drug based on the specification information obtained from Drugbank. They compute the Jaccard similarity of the fingerprints. A Jaccard similarity score between two sets X and Y is defined as:
\begin{center}
$Jaccard(X,Y)=\frac{|X\bigcap Y|}{|X\bigcup Y|}$
\end{center}
\item \textit{Ligand-based drug similarity:} \cite{perlman2011combining} compare the specification information from Drugbank against a collection of ligand sets using the similarity ensemble approach (SEA) search tool \cite{keiser2009predicting}. The ligand sets denote to the Jaccard similarity between the corresponding sets of protein-receptor families for each drug pair.
\item \textit{Expression-based drug similarity:} \cite{perlman2011combining} use the Spearman rank correlation coefficient to compute a similarity of how gene expression responses to drugs which is obtained from the Connectivity Map Project \cite{lamb2007connectivity,lamb2006connectivity}. The Spearman rank correlation coefficient between two sets X and Y is calculated as:
\begin{center}
$Spearman(X,Y)=\frac{\sum_i(x_i-\overline x)(y_i-\overline y)}{\sqrt{\sum_i(x_i-\overline x)^2\sum_i(y_i-\overline y)^2}}$
\end{center}
\item \textit{Side-effect-based drug similarity:} The Jaccard similarity of side-effect sets for each drug pairs. The side-effect sets are obtained from \cite{gunther2007supertarget}.
\item \textit{Annotation-based drug similarity:} \cite{perlman2011combining} use the semantic similarity algorithm of Resnik (Resnik, 1999) to calculate the similarity of ATC code which is obtained from DrugBank and matched against the World Health Organization ATC classification system \cite{skrbo2004classification} for each drug pair.
\item \textit{Sequence-based target similarity:} \cite{perlman2011combining} use the sequence-based similarity score as a target similarity measure which is suggested in \cite{bleakley2009supervised}.
\item \textit{Protein-protein interaction network-based target similarity:} They calculate the distance between target pairs as similarity measure using an all-pairs shortest path algorithm on the human protein-protein interactions network.
\item \textit{Gene Ontology-based target similarity:} They compute the semantic similarity between Gene Ontology annotations from the source of \cite{jain2009infrastructure} using Resnik's method \cite{resnik1999semantic}
\end{itemize}

We follow the same procedure as \cite{6817596} that uses a ten-folder validation for the experiments. Each folder we have $90\%$ of positive links and negative links for training, and the remaining $10\%$ of links for testing.

The pre-defined rules for this multi-similarity dataset is showned below:

\begin{flushleft}

{$m_1: D1 \ra\limits^{Chemical} D2 \ra\limits^{Interacts} T1$

$m_2: D1 \ra\limits^{Ligand} D2 \ra\limits^{Interacts} T1$

$m_3: D1 \ra\limits^{Expression} D2 \ra\limits^{Interacts} T1$

$m_4: D1 \ra\limits^{Side-effect} D2 \ra\limits^{Interacts} T1$

$m_5: D1 \ra\limits^{Annotation} D2 \ra\limits^{Interacts} T1$

$m_6: D1 \ra\limits^{Interacts} T2 \ra\limits^{Sequence} T1$

$m_7: D1 \ra\limits^{Interacts} T2 \ra\limits^{Protein} T1$

$m_8: D1 \ra\limits^{Interacts} T2 \ra\limits^{Gene Ontology} T1$}

\end{flushleft}

\subsubsection{Dataset II}
A total of up to nine types of nodes are presented, including drugs (compounds), targets (proteins), adverse side effects, Gene Ontology (GO) annotations, ChEBI types, substructures, tissues, biological pathways and diseases; ten types of edges are presented, including drug-ChEBI types associations, drug-protein interactions, drug-substructure associations, adverse side effect-drug associations, disease-drug associations, target-target interactions, target-GO annotation associations, disease-protein associations, pathway-protein associations and tissue-target associations. In addition, a 2D structural based drug-drug similarity measure and a sequence based target-target similarity measure are included. The similarity measures are obtained from the PubChem databases \cite{kim2015pubchem,wang2009pubchem}. The total number of nodes in this heterogeneous network is 295897, including 258030 drugs and 22056 targets, and the total number of edges in the network is 7191240, the total number of meta-paths is 6487339992.

We follow the procedure in \cite{Fu2016} that a set of 145,622 positively labeled DTI links and 600,000 negatively labeled DTI links contain in the current heterogeneous network are treated as the training set; another set of 43,159 positive links and 195,000 negative links that are not observed in the network are treated as the testing set.

We follow the method of \cite{Fu2016} that defines 51 different semantic meta-paths on the heterogeneous network. A sample of meta-paths is presented in Table 1, and the full version of 51 meta-paths is shown in the supplement.

\begin{table}
  \centering
  \caption{Sample of Semantic meta-paths defined for Dataset II}
  \label{tab:table1}
  \begin{tabular}{cl}
    \toprule
    Index & Semantics\\
    \midrule
    C1 & $drug \ra\limits^{similar\ to} drug \ra\limits^{interacts} target$\\
    C2 & $drug \ra\limits^{interacts} target \ra\limits^{interacts} target$\\
    C3 & $drug \ra\limits^{interacts} target \ra\limits^{similar\ to} target$\\
    C4 & $drug \ra\limits^{treats} disease \ra\limits^{caused\ by} target$\\
    C5 & $drug \ra\limits^{similar\ to} drug \ra\limits^{interacts} drug \ra\limits^{interacts} target$\\
    C11 & $drug \ra\limits^{treats} disease  \ra\limits^{treated\ by} drug \ra\limits^{interacts} target$\\
    C14 & $drug \ra\limits^{interacts} target \ra\limits^{expressed\ in} tissue \ra\limits^{expresses} target$\\
    C15 & $drug \ra\limits^{interacts} target  \ra\limits^{causes} disease \ra\limits^{caused\ by} target$\\
    \bottomrule
  \end{tabular}
\end{table}

\subsubsection{Dataset III}
A total of four types of nodes (drugs, proteins, diseases and side-effects) and six types of edges (drug-protein interactions, drug-drug interactions, drug-disease associations, drug-side-effect associations, protein-disease associations and protein-protein interactions) representing diverse drug-related information are collected from the public databases that were used to construct this dataset \cite{luo2017network}. It contains 12,015 nodes and 1,895,445 edges in total, including 708 drugs and 1512 drugs. A set of drug and target similarity measures is also included.

The known DTIs as well as drug-drug interactions are collected from DrugBank \cite{wishart2007drugbank}, the protein-protein interactions are collected from the HPRD database \cite{keshava2008human}, the drug-disease and protein-disease associations are collected from the Comparative Toxicogenomics Datasbase \cite{davis2012comparative} and the drug-side-effect associations are collected from the SIDER database \cite{kuhn2010side}.

We follow the same experimental settings in \cite{luo2017network} that randomly pick $90\%$ of positive DTI links and corresponding number of negative links as training set, and the remaining $10\%$ positive links and corresponding number of negative links as testing set. Since the information type provided in Datset III is a subset of Dataset II, we pick a subset of 21 meta-paths out of 51 used in Dataset II. The subset meta-path we select is: \{C1, C2, C3, C4, C5, C6, C7, C11, C15, C16, C17, C18, C19, C24, C25, C26, C27, C44, C45, C46, C47\}.

\subsection{Evaluation Metrics}

We use the area under the Precision-Recall curve - the AUPR score and the area under the Receiver Operatin Characteristic (ROC) curve - the AUC score as measurements.

The AUC score is a commonly used measurement of a binary classifier in related pubilications. The ROC curve is plotted by the true positive rate (TPR) against the false-positive rate (FPR) at various thresholds. By applying the AUC score, we can compare our model with many other papers' approaches. However, if a dataset is highly imbalanced \textit{i.e.} the number of positive test cases is too small compared with the number of negative test cases, the change of AUC score will be subtle. So we propose a second evaluation criteria, the AUPR score, since it can be more informative than the AUC score under an imbalanced dataset.

\subsection{Baselines}

We compare our model with five approaches: original PSL method \cite{6817596}, Meta-path count feature + Random Forests and Support Vector Machine methods \cite{Fu2016}, the Pairwise MKL method \cite{cichonska2018learning} and the DTINet method \cite{luo2017network}. All five baselines can predict drug-target interactions on a heterogeneous network. 

\subsubsection{Probabilistic Soft Logic}

\cite{6817596} introduces the Probabilistic Soft Logic model, which pre-defines a series of association rules, treated as "rules", and solves the DTI prediction problem as inference on a bipartite graph. More specifically, it introduces eight different similarity based association rules for the DTI prediction task. (Shown at section 6.1.1) After defining the rules, \cite{6817596} incorporates all rule instances within a bipartite graph, and minimizes a total \textit{distance to satisfaction} based on all rule instances to make predictions.

\cite{Fu2016} introduces the Meta-path topological feature, which also defines a series of association rules but on a heterogeneous network, denoted as "Meta-path", then uses matrix multiplications to calculate a Meta-path count. For the DTI prediction task, they define 51 different meta-paths (introduced in section 6.1.2) and for each drug-target pair, they form a 51-dimensional vector based on the meta-path count topological features, then uses machine learning classifiers, Support Vector Machine and Random Forests, to solve the DTI prediction problem as a supervised learning task.

Moreover, \cite{luo2017network} introduces a method that could generate a low dimensional representation for both drug and target, based on a series of association matrices. For each drug, they have four association matrices and for each target, they have three association matrices. They apply a diffusion component analysis (DCA), to obtain the informative low-dimensional feature representations, then use a learned projection matrix that could project the drug feature into protein space so that the distance between the projected vectors and the interacted targets are minimized.

We also compare our model with a latest approach \cite{cichonska2018learning} that uses a kernel-based approach that learns a set of weights for each single kernel as well as for each combination of kernel pairs.

\subsection{Results}
We operate several experiments including both effectiveness comparison and running time comparison, and we compare results between different approaches on all three datasets introduced in section 6.1. Moreover, we report a case study of the weight learning and the selection of meta-paths. 

\subsubsection{Effectiveness Comparison}

Table 2 shows the comparison experimental results between our model and other approaches on all datasets in AUC score, and AUPR score.

\begin{table*}[h!]
  \centering
  \caption{Effectiveness comparison between our model and baselines on all datasets}
  \label{tab:table1}
  \begin{tabular}{ccccccc}
    \toprule
     & PSL triads & Meta-path+SVM & Meta-path+RFs & DTINet & PairwiseMKL & SMPSL\\
    \midrule
    AUC in Dataset I & 0.920 & 0.719 & 0.766 & 0.844 & 0.825 & \textbf{0.929}\\
    AUPR in Dataset I & \textbf{0.617} & 0.378 & 0.430 & 0.381 & 0.225 & \textbf{0.617}\\
    \midrule
    AUC in Dataset II & N/A & 0.867$^*$ & 0.845 \& 0.542 (self) & N/A & N/A & \textbf{0.917}\\
    AUPR in Dataset II & N/A & 0.523$^*$ & N/A \& 0.248 (self) & N/A & N/A & \textbf{0.815}\\
    \midrule
    AUC in Dataset III & N/A & 0.509$^*$ & 0.884 & 0.914 & N/A & \textbf{0.928}\\
    AUPR in Dataset III & N/A & 0.505$^*$ & 0.884 & 0.932 & N/A & \textbf{0.947}\\
    \bottomrule
  \end{tabular}
\end{table*}

From the experimental results of Dataset I we can tell that the PSL approach proposed in 2014 still outperforms the Metapath+SVM/RFs (2016), DTINet (2017) and PairwiseMKL (2018) methods. Furthermore, our Summated Meta-path PSL model preserves the performance after applying the summation and gives a slightly better results in AUC score. 

The Dataset II is a very huge dataset: the total number of meta-path counts is 6487339992, over 100000 times the size of Dataset I (which is 38251); the number of drugs and targets are also hundreds times the size of Dataset I. As a result, the PSL, DTINet and PairwiseMKL model fail to finish running in a reasonable time limit (168 hours), because they do not apply any summation strategies. For the Meta-path + SVM method, we only sampled 1\% of training data to feed the classifier. The AUC score (0.845) of the Meta-path + RFs method is proposed by the original paper \cite{Fu2016}, yet the AUPR score are not reported. Based on the reported parameters in the original paper, we implement a Python based CM-RFs model using the sklearn.RandomForestClassifier() function, and the parameters are: $n\_estimators = 500, max\_features = 13$.

In terms of Dataset III, we apply the same experiment protocol as \cite{luo2017network} that runs a ten-folder cross-validation, each folder randomly picks 90\% of positive links and corresponding number of negative links as training, and the remaining 10\% of positive links corresponding with the same number of randomly picked negative links as the test set. By doing so, each validation folder only contains 3460 labeled drug-target interaction links and 384 links to be predicted, which helps the DTINet approach executable. Yet the total number of meta-path counts is still very large, which is 783950268303, because there are multiple association and similarity measures on the heterogeneous network. As a result, the PSL model and the PairwiseMKL model still cannot finish running on this dataset in a reasonable amount of time.

\subsubsection{Running Time Comparison}

More importantly, due to the fact that we use a summation strategy to reduce the number of rule instances of a PSL model, we can accelerate the DTI prediction process.  Table 3 shows the running time comparison between all five approaches on three datasets.

\begin{table*}[htbp]
  \centering
  \caption{Running time comparison between our model and baselines on all datasets (in minutes)}
  \label{tab:table1}
  \begin{tabular}{cccc}
    \toprule
    Methods & $Dataset\ I$ & $Dataset\ II$ & $Dataset\ III$\\
    \midrule
    PSL triads & 14.95 & $>10000$ & $>10000$\\
    Meta-path+SVM & 9.32 & 3.68$^*$ & 3.85$^*$\\
    Meta-path+RFs & 1.32 & N/A \& 32.53 (self) & 0.913\\
    DTINet & 19.13 & $>10000$ & 18.83\\
    PairwiseMKL & 386 & $>10000$ & $>10000$\\
    \midrule    
    SMPSL & \textbf{0.159} & \textbf{2.03} & \textbf{0.85}\\
    \bottomrule
  \end{tabular}
\end{table*}

From the dataset I result we can tell that the pairwiseMKL method takes the moust amount of time. In the original paper \cite{cichonska2018learning}, the experiments were done on a 120 drug 120 cancer bipartite network, and the running time was 1.45h. Thus, the running time for pairwiseMKL on dataset I is logical.

We run our experiments on a computer with a $(16 \times 2)$ 3.3 GHz Intel Xeon CPU and 128GB of RAM. We gain significant efficiency improvement compared with the original PSL model, resulting in over 99\% of time reduction while gaining comparable results. Even comparing with other approaches, our method still takes the least amount of time.

\subsubsection{Case Study: Weight Learning}

We study the effect of weight learning. The effectiveness and running time comparison with/without weight learning is shown below:

\begin{table}[h!]
  \centering
  \caption{Effectiveness and running time comparison with/without weight learning on dataset I, II, III acccordingly}
  \label{tab:table1}
  \begin{tabular}{lccc}
    \toprule
    Methods & AUC & AUPR & Running Time(min)\\
    \midrule
    (I)weight learning & 0.929 & 0.617 & $0.159$\\
    (I)no weight learning & 0.927 & 0.0.552 &0.142\\
    (II)weight learning & 0.917 & 0.815 & $2.030$\\
    (II)no weight learning & 0.911 & 0.779 & 1.815\\
    (III)weight learning & 0.928 & 0.947 & $0.850$\\
    (III)no weight learning & 0.926 & 0.945 & 0.814\\
    \bottomrule
  \end{tabular}
\end{table}

From Table 4 we can tell that weight learning is positively effective, especially when the dataset is imbalanced (For Dataset I, the ratio of positive links and negative links in the test set is approximately 1 : 45; For Dataset II, the ratio is 1 : 4.5; For Dataset III, the ratio is 1 : 1).

We also demonstrate a figure that shows the log-scale importance of the relative weight parameter $w$ in Dataset II:

\begin{figure}[h!]
\centering\includegraphics[width=3.2in]{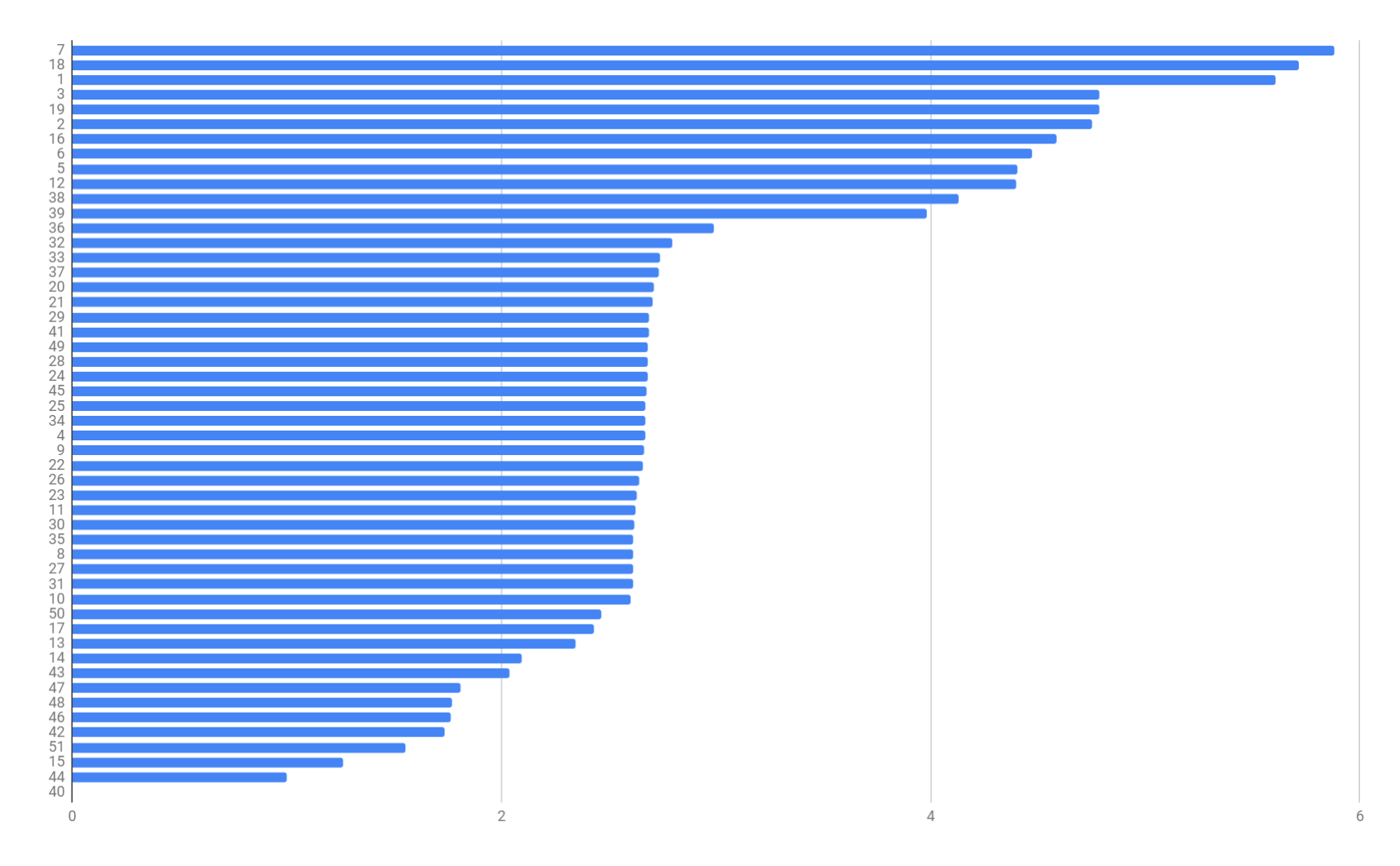}
\caption{The log-scale importance of the relative weight parameter $w$ in Dataset II}\label{Figure 1} 
\end{figure}

The weight parameter $w$ indicates the importance of a meta-path. From figure 2 we can tell that the top-6 most effective meta-paths selected by our model is: $\{C7, C18, C1, C3, C19, C2\}$. We can draw a conclusion from this case study that shorter meta-paths ${C1, C2, C3}$ tend to have higher effect. Besides, all the top-6 meta-paths are defined using only interaction and similarity association measures within drugs and targets. Which satisfies the truth that interaction and similarity information are more solid in the DTI prediction task.

\section{Conclusion}
In this paper, we propose a summated meta-path and probabilistic soft logic model (SMPSL) for the drug-target interaction (DTI) prediction. We form the DTI prediction problem into an inference problem on a heterogeneous network with multiple domain-specific measures, such as similarities, associations, interactions etc. We detect the shortcut for both Meta-path method and PSL method, while succeed in combining both method together, showing success to transfer the meta-path topological feature into a probabilistic metric by generating the probabilistic commuting matrices based on a Bayesian probabilistic approach. By using the value of commuting matrices as the rule instance of the PSL model, we significantly reduce the total number of rule instances of a PSL model, drawing over 99\% of time reductions while the performance still remains comparable in AUC score and AUPR score. Besides, we compare our model with other four latest DTI prediction approaches on three open-source large-scale datasets, showing performance improvement in both AUC/AUPR score and framework efficiency.

In addition, it is worth pointing out that our method may be eligible to extend to broader scope of applications because both the meta-path method and the PSL method are general tools widely used in various areas, including but not constrained in spammer detection, scheme mapping, page ranking, recommendation system \textit{etc.} This possibility motivates us to keep digging the potential of our model.

%
\bibliographystyle{ACM-Reference-Format}
\bibliography{acmart}

%
\appendix

\section{Supplement}

Table 4 and Table 5 show the total 51 meta-paths defined for Dataset II:

\begin{table*}
  \centering
  \caption{Semantic meta-paths defined for Dataset II Part A }
  \label{tab:table1}
  \begin{tabular}{cl}
    \toprule
    Index & Semantics\\
    \midrule
    C1 & $drug \ra\limits^{similar\ to} drug \ra\limits^{interacts} target$\\
    C2 & $drug \ra\limits^{interacts} target \ra\limits^{interacts} target$\\
    C3 & $drug \ra\limits^{interacts} target \ra\limits^{similar\ to} target$\\
    C4 & $drug \ra\limits^{treats} disease \ra\limits^{caused\ by} target$\\
    C5 & $drug \ra\limits^{similar\ to} drug \ra\limits^{interacts} drug \ra\limits^{interacts} target$\\
    C6 & $drug \ra\limits^{similar\ to} drug \ra\limits^{interacts} target\ra\limits^{interacts} target$\\
    C7 & $drug \ra\limits^{interacts} target \ra\limits^{interacts} drug \ra\limits^{interacts} target$\\
    C8 & $drug \ra\limits^{has\ part} substructure \ra\limits^{part\ of} drug \ra\limits^{interacts} target$\\
    C9 & $drug \ra\limits^{has\ type} ChEBI\ type \ra\limits^{type\ of} drug \ra\limits^{interacts} target$\\
    C10 & $drug \ra\limits^{induces} adverse\ side\ effect \ra\limits^{induced\ by} drug \ra\limits^{interacts} target$\\
    C11 & $drug \ra\limits^{treats} disease  \ra\limits^{treated\ by} drug \ra\limits^{interacts} target$\\
    C12 & $drug \ra\limits^{interacts} target \ra\limits^{has\ annotation} GO\ annotation \ra\limits^{annotation\ of} target$\\
    C13 & $drug \ra\limits^{interacts} target  \ra\limits^{participates\ in} pathway \ra\limits^{has\ participants} target$\\
    C14 & $drug \ra\limits^{interacts} target \ra\limits^{expressed\ in} tissue \ra\limits^{expresses} target$\\
    C15 & $drug \ra\limits^{interacts} target  \ra\limits^{causes} disease \ra\limits^{caused\ by} target$\\
    C16 & $drug \ra\limits^{interacts} target \ra\limits^{interacts} drug \ra\limits^{interacts} target \ra\limits^{interacts} target$\\
    C17 & $drug \ra\limits^{interacts} target \ra\limits^{interacts} drug \ra\limits^{treats} disease \ra\limits^{caused\ by} target$\\
    C18 & $drug \ra\limits^{interacts} target \ra\limits^{interacts} drug \ra\limits^{interacts} drug \ra\limits^{interacts} target$\\
    C19 & $drug \ra\limits^{interacts} target \ra\limits^{interacts} drug \ra\limits^{interacts} target \ra\limits^{similar\ to} target$\\
    C20 & $drug \ra\limits^{type\ of} ChEBI\ type \ra\limits^{type\ of} drug \ra\limits^{interacts} target \ra\limits^{interacts} target$\\
    C21 & $drug \ra\limits^{type\ of} ChEBI\ type \ra\limits^{type\ of} drug \ra\limits^{treats} disease \ra\limits^{caused\ by} target$\\
    C22 & $drug \ra\limits^{type\ of} ChEBI\ type \ra\limits^{type\ of} drug \ra\limits^{interacts} target \ra\limits^{similar\ to} target$\\
    C23 & $drug \ra\limits^{type\ of} ChEBI\ type \ra\limits^{type\ of} drug \ra\limits^{similar\ to} drug \ra\limits^{interacts} target$\\
    C24 & $drug \ra\limits^{treats} disease \ra\limits^{treated\ by} drug \ra\limits^{interacts} target \ra\limits^{interacts} target$\\
    C25 & $drug \ra\limits^{treats} disease \ra\limits^{treated\ by} drug \ra\limits^{treats} disease \ra\limits^{caused\ by} target$\\
    C26 & $drug \ra\limits^{treats} disease \ra\limits^{treated\ by} drug \ra\limits^{interacts} target \ra\limits^{similar\ to} target$\\
    C27 & $drug \ra\limits^{treats} disease \ra\limits^{treated\ by} drug \ra\limits^{similar\ to} drug \ra\limits^{interacts} target$\\
    C28 & $drug \ra\limits^{induces} adverse\ side\ effect \ra\limits^{induced\ by} drug \ra\limits^{interacts} target \ra\limits^{interacts} target$\\
    C29 & $drug \ra\limits^{induces} adverse\ side\ effect \ra\limits^{induced\ by} drug \ra\limits^{treats} disease \ra\limits^{caused\ by} target$\\
    C30 & $drug \ra\limits^{induces} adverse\ side\ effect \ra\limits^{induced\ by} drug \ra\limits^{interacts} target \ra\limits^{similar\ to} target$\\
    C31 & $drug \ra\limits^{induces} adverse\ side\ effect \ra\limits^{induced\ by} drug \ra\limits^{similar\ to} drug \ra\limits^{interacts} target$\\
    C32 & $drug \ra\limits^{has\ part} substructure \ra\limits^{part\ of} drug \ra\limits^{interacts} target \ra\limits^{interacts} target$\\
    C33 & $drug \ra\limits^{has\ part} substructure \ra\limits^{part\ of} drug \ra\limits^{treats} disease \ra\limits^{caused\ by} target$\\
    C34 & $drug \ra\limits^{has\ part} substructure \ra\limits^{part\ of} drug \ra\limits^{interacts} target \ra\limits^{similar\ to} target$\\
    C35 & $drug \ra\limits^{has\ part} substructure \ra\limits^{part\ of} drug \ra\limits^{similar\ to} drug \ra\limits^{interacts} target$\\
    C36 & $drug \ra\limits^{interacts} target \ra\limits^{interacts} target \ra\limits^{has\ annotation} GO\ annotation \ra\limits^{annotation\ of} target$\\
    \bottomrule
  \end{tabular}
\end{table*}
\begin{table*}
  \centering
  \caption{Semantic meta-paths defined for Dataset II part B}
  \label{tab:table1}
  \begin{tabular}{cl}
    \toprule
    Index & Semantics\\
    \midrule
    C37 & $drug \ra\limits^{treats} disease \ra\limits^{caused\ by} target \ra\limits^{has\ annotation} GO\ annotation \ra\limits^{annotation\ of} target$\\
    C38 & $drug \ra\limits^{interacts} target \ra\limits^{similar\ to} target \ra\limits^{has\ annotation} GO\ annotation \ra\limits^{annotation\ of} target$\\
    C39 & $drug \ra\limits^{similar\ to} drug \ra\limits^{interacts} target \ra\limits^{has\ annotation} GO\ annotation \ra\limits^{annotation\ of} target$\\
    C40 & $drug \ra\limits^{interacts} target \ra\limits^{interacts} target \ra\limits^{participates\ in} pathway \ra\limits^{has\ participants} target$\\
    C41 & $drug \ra\limits^{treats} disease \ra\limits^{caused\ by} target \ra\limits^{participates\ in} pathway \ra\limits^{has\ participants} target$\\
    C42 & $drug \ra\limits^{interacts} target \ra\limits^{similar\ to} target \ra\limits^{participates\ in} pathway \ra\limits^{has\ participants} target$\\
    C43 & $drug \ra\limits^{similar\ to} drug \ra\limits^{interacts} target \ra\limits^{participates\ in} pathway \ra\limits^{has\ participants} target$\\
    C44 & $drug \ra\limits^{interacts} target \ra\limits^{interacts} target \ra\limits^{causes} disease \ra\limits^{caused\ by} target$\\
    C45 & $drug \ra\limits^{treats} disease \ra\limits^{caused\ by} target \ra\limits^{causes} disease \ra\limits^{caused\ by} target$\\
    C46 & $drug \ra\limits^{interacts} target \ra\limits^{similar\ to} target \ra\limits^{causes} disease \ra\limits^{caused\ by} target$\\
    C47 & $drug \ra\limits^{similar\ to} drug \ra\limits^{interacts} target \ra\limits^{causes} disease \ra\limits^{caused\ by} target$\\
    C48 & $drug \ra\limits^{interacts} target \ra\limits^{interacts} target \ra\limits^{expressed\ in} tissue \ra\limits^{expresses} target$\\
    C49 & $drug \ra\limits^{treats} disease \ra\limits^{caused\ by} target \ra\limits^{expressed\ in} tissue \ra\limits^{expresses} target$\\
    C50 & $drug \ra\limits^{interacts} target \ra\limits^{similar\ to} target \ra\limits^{expressed\ in} tissue \ra\limits^{expresses} target$\\
    C51 & $drug \ra\limits^{similar\ to} drug \ra\limits^{interacts} target \ra\limits^{expressed\ in} tissue \ra\limits^{expresses} target$\\
\bottomrule
  \end{tabular}
\end{table*}

\end{document}